\documentclass[a4paper,UKenglish]{lipics}
 
\usepackage{microtype}

\usepackage{xcolor}
\usepackage{pict2e}

\title{Logic Considered Fun}
\author{John Slaney}
\affil{Australian National University \\
  \texttt{John.Slaney@anu.edu.au}}
\authorrunning{J. Slaney} 

\Copyright{John Slaney}


\serieslogo{logo_ttl}
\volumeinfo
  {M. Antonia {Huertas}, Jo\~ao {Marcos}, Mar\'ia {Manzano}, Sophie {Pinchinat}, \\
  Fran\c{c}ois {Schwarzentruber}}
  {5}
  {4th International Conference on Tools for Teaching Logic}
  {1}
  {1}
  {215}
\EventShortName{TTL2015}

\newcommand{\Indent}[1]{\hspace*{#1em}}
\newcommand{\connective}[2]{\makebox[#1em]{#2}}
\newcommand{\K}{\connective{1}{$\wedge$}}
\newcommand{\C}{\connective{1.2}{$\rightarrow$}}

\begin{document}

\maketitle

\begin{abstract}
This report describes the development and use of an online teaching tool giving students exercises in logical modelling, or \emph{formalisation} as it is called in the older literature. The original version of the site, `Logic for Fun', dates from 2001, though it was little used except by small groups of students at the Australian National University. It is currently in the process of being replaced by a new version, free to all Internet users, intended to be promoted widely as a useful addition to both online and traditional logic courses.
\end{abstract}

\section{Background: Formalisation}

In teaching formal logic to undergraduates, we attempt to impart a range of skills. In a typical ``Logic 101'' course, the most prominent of these involve manipulation of calculi: devising proofs, usually using some form of natural deduction, constructing semantic tableaux or the like. We also ask students to formalise natural language sentences---often specially constructed to involve awkward nesting of connectives or strings of quantifiers---and may hope that they acquire some facility in critical reasoning and perhaps an appreciation of some wider issues connected to logic, be they mathematical, computational, philosophical or historical. Some of these things we teach better than others. Most students do become tolerably adept at handling the technical details of natural deduction proofs, but in many cases they remain depressingly unable to write a well-formed formula to express even a simple claim about a domain of discourse.%
\footnote{Evidence for this claim is anecdotal, but strong. My own appreciation of it was sharpened in 2011, when analysis of results from a class of 61 students showed grades on proof construction that were on average above their grades for other courses, but after 13 weeks of study more than a third of them were unable to express `The bigger the burger the better the burger' adequately in the notation of first order logic.}
 Barwise and Etchemendy, for instance, comment:
\begin{quote}
The real problem, as we see it, is a failure on the part of logicians to find a
simple way to explain the relationship between meaning and the laws of logic.
In particular, we do not succeed in conveying to students what sentences
in FOL mean, or in conveying how the meanings of sentences govern which
methods of inference are valid and which are not. \hfill \cite{barwise91}, p.13
\end{quote}
We should find this situation alarming. Mechanical symbol-pushing for the purposes of simple proofs is easy to teach, fairly easy to learn and almost useless once the course is finished. On the other hand, the ability to read and write in the notation of formal logic, to use this as a medium for knowledge representation, to analyse and to disambiguate, is the most important skill students can take away from an introductory logic course, and it is a skill most of us can claim less success in teaching.

It was against this background that the tool \emph{Logic for Fun} \cite{LFFold,LFFnew} was devised around 15 years ago. \emph{Logic for Fun} is a website on which users are invited to express a range of logical problems and puzzles in such a way that a black-box solver on the site can produce solutions. The language in which problems are to be expressed is that of a many-sorted first order logic, extended slightly with a few built-in expressions and modest support for integers. The solver takes as input a set of formulae in this language and searches for finite models of this set. If it finds a unique model, this is almost certainly a solution to the problem. More usually it either reports syntax errors in the encoding or else says that there is no solution. The user (i.e.\ the student) then debugs the encoding until it is correct. This has several advantages over traditional formalisation exercises:
\begin{enumerate}
\item The sentences to be formalised constitute a meaningful problem, rather than looking like isolated examples of things it may be tricky to express formally;
\item Feedback is always accurate and \emph{immediate}, rather than coming a week after the homework was handed in;
\item Because there is no feedback latency, because there is a goal (solving the problem) and because the machine is infinitely patient, students will put time and effort into their work, to a degree never seen in the traditional setting.
\end{enumerate}

\begin{figure}\centering
\includegraphics[width=120mm]{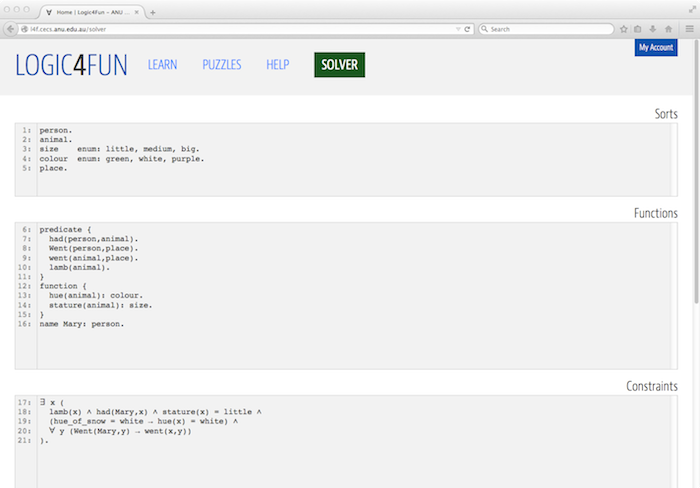}
\caption{\label{fig:L4Fwindow}Screenshot of the window for problem input}
\end{figure}

An example (not a puzzle in this case, but a first order theory) will help to illustrate the process required of the student. Figure \ref{fig:L4Fwindow} shows the form used for text input. It consists of three boxes: in the first are listed the ``sorts'' or domains over which variables are to range; in the second is the vocabulary of non-logical symbols (predicates, names, etc) with their types; in the third are the constraints, written as first order formulae. There are also buttons, not visible in the screenshot,  for running the solver---one for generating solutions and another for running in a lightweight mode to check syntax. In the example, there are 5 sorts (persons, animals, sizes, colours and places) of which two (sizes and colours) are given by explicit enumeration while the other three are left for the solver to decide.

\noindent{\tt \mbox{} \\
 \Indent{1} Sorts: \\
 \Indent{2} person. \\
 \Indent{2} animal. \\
 \Indent{2} size enum: little, medium, big. \\
 \Indent{2} colour enum: green, white, purple. \\
 \Indent{2} place. \\ \mbox{}
}

\noindent
The domains of these sorts are required to be disjoint. We wish to say that Mary had a little lamb, so as vocabulary we declare `Mary' as a name, `had' as a relation between persons and animals, `stature' as a function of animals and `lamb' as a predicate picking out a subset of the animals:

\noindent{\tt \mbox{} \\
 \Indent{1} Vocabulary: \\
 \Indent{2} predicate $\{$ \\
 \Indent{3} had(person,animal). \\
 \Indent{3} Went(person,place). \\
 \Indent{3} went(animal,place). \ \ \ \% note: case-sensitive \\
 \Indent{3} lamb(animal). \\
 \Indent{2} $\}$ \\
 \Indent{2} function $\{$ \\
 \Indent{3} hue(animal): colour. \\
 \Indent{3} stature(animal): size. \\
 \Indent{2} $\}$ \\
 \Indent{2} name Mary: person. \\ \mbox{}
}

\noindent
For knowledge representation purposes, it is very convenient to use a many-sorted logic and to specify the types of predicates and function symbols separately from the formulae in which they occur. The hue and stature functions, for example, map animals to their colours and sizes; to model the constraints will be to determine which functions to assign to them.

To finish the example, here is the constraint:

\noindent{\tt \mbox{} \\
 \Indent{1} $\exists$ x (had(Mary,x) \K\ stature(x) = little \K\ lamb(x) \K \\
 \Indent{3} (hue\_of\_snow = white \C\ hue(x) = white) \K \\
 \Indent{3} $\forall$ y (Went(Mary,y) \C\ went(x,y))). \\ \mbox{}
}

\noindent
Having written the description, the student clicks ``Solve'' and the back-end reasoner (essentially a SAT solver adapted to finite domain first order problems) starts searching for solutions. Naturally, it finds models of this little theory with \emph{very} small domains, including the expected interpretation in which there is one person (Mary), one animal (the little white lamb) and one place which neither of them visits. Amusingly, the solver also finds unexpected solutions, for instance in which the lamb is green---but it is still just as white as snow because the hue of snow is purple! In the pedagogic context, this provides a good opportunity for the teacher to make some points about interpretations and truth conditions and the semantics of the material conditional. When this kind of situation arises in modelling a puzzle which should admit only one solution, the student must invent more constraints to supply the missing information (e.g.\ that snow is white and that Mary went to school). In order to do this, they have to think about the semantics of the problem, render it into first order logic and understand the relationship between the formulae they have written and the satisfying formal interpretations.

\section{Structure of the exercises}

The problems given as exercises on the site are divided into five levels: Beginner, Intermediate, Advanced, Expert and Logician. The boundaries between levels are not really definite, but students like the idea of progressing through levels in the style of a game. ``Beginner'' problems are fairly trivial to represent and solve, and are designed to get students through the phase of learning to use the site, teaching them how to declare vocabulary and the like. ``Intermediate'' ones are mainly logic puzzles of the kind found in popular magazines, often calling for bijections between sets of five or so things satisfying a list of clues. ``Advanced'' puzzles are not necessarily harder, but have features requiring more sophisticated logical treatment---nested quantifiers and the like. The ``Expert'' puzzles are more challenging, and include several state-transition problems from AI planning, for instance. They require students to supply less obvious vocabulary and axiomatisation to represent preconditions, postconditions and frame conditions of actions and the like. Finally, the ``Logician'' section contains problems which hint at applications of logic, for instance to finite combinatorics and to model-based diagnosis.

It is important that all of the suggested problems can be solved quickly by the software behind the site, without requiring coding tricks. This is because the aim is to teach correct logical expression, not constraint programming. Especially for some of the ``logician'' puzzles, efficiency does matter, as the underlying problems are NP-hard at best, and solver behaviour can be affected by non-obvious things like the order in which functions are declared, but as far as possible the site de-emphasises efficiency and instead lays stress on correctness.

An interesting feature of learning logical modelling in this way is that concepts are introduced in approximately the opposite order from that in the parallel lecture course. In a typical logic course, we work from the more abstract levels via propositional connectives, then to names, predicates, quantifier-variable notation, and then introduce identity as a special relation symbol and go on finally to deal with function symbols and general terms. We do not usually get as far as many-sorted logic. Logical modelling \emph{starts} with sorts, equations, names and functions. Then generality is introduced (the universal quantifier is used much more than the existential one), and only after that the basic connectives. Introductory logic textbooks, though they vary greatly in emphasis and style, tend to follow the same overall direction as logic courses.%
\footnote{There are exceptions. One of the most notable is the little introduction by Wilfred Hodges \cite{hodges} used over 30 years ago as a textbook by the Open University and still in print.}
My experience of using \emph{Logic for Fun} as part of a standard introductory logic course is that the reversed order of topics requires that the lecturer put work into pointing out the relationship between the logical modelling exercises and the rest of the course, as the two strands do not converge until the end.

\section{Errors and feedback}

\setlength{\unitlength}{6mm}

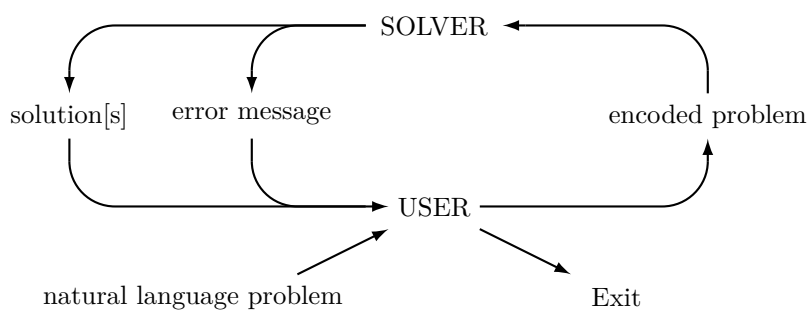
\begin{figure}
\centering
\begin{picture}(0,7)(0,-1)
\thicklines
\put(0,2){\makebox(0,0){USER}}
\put(0,6){\makebox(0,0){SOLVER}}
\put(-8,4){\makebox(0,0){solution[s]}}
\put(-4,4){\makebox(0,0){error message}}
\put(6,4){\makebox(0,0){encoded problem}}
\put(-2,0){\makebox(0,0)[r]{natural language problem}}
\put(4,0){\makebox(0,0){Exit}}

\put(-3,0.5){\vector(2,1){2}}
\put(1,1.5){\vector(2,-1){2}}
\put(1,3){\oval(10,2)[br]}
\put(6,3){\vector(0,1){0.5}}
\put(6,4.5){\line(0,1){0.5}}
\put(2,5){\oval(8,2)[tr]}
\put(2,6){\vector(-1,0){0.5}}
\put(-1.5,5){\oval(5,2)[tl]}
\put(-4,5){\vector(0,-1){0.5}}
\put(-1.5,5){\oval(13,2)[tl]}
\put(-8,5){\vector(0,-1){0.5}}
\put(-4,3.5){\line(0,-1){0.5}}
\put(-1.5,3){\oval(5,2)[bl]}
\put(-1.5,2){\vector(1,0){0.5}}
\put(-8,3.5){\line(0,-1){0.5}}
\put(-1.5,3){\oval(13,2)[bl]}
\end{picture}
\caption{\label{fig:workflow}\emph{Logic for Fun} workflow. The main loop is a dialogue between a user and the solver.}
\end{figure}

The workflow of the site is one of dialogue between student and machine, whereby the problem in natural language is initially proposed as a challenge to which the student responds by writing formal encodings of all or part of the problem which the solver evaluates. Feedback in the form of error messages or solutions (or lack of solutions) informs the student's next attempt. The cycle is broken when the student decides to terminate or suspend it. Work may be saved at any point for future reference.

Clearly, the educationally effective part of this process is the correction of errors. To put it simply: the tool is only doing useful work when its users are making mistakes. Feedback is therefore the essence of the process. Errors (apart from accidental slips) are fundamentally of two kinds, syntactic or semantic, evoking very different responses from the system. Errors of syntax are caught by the parser or the type checker and reported with explicit messages. For example, if the user writes \\[1ex]
{\tt
  \Indent{2} had(Mary, $\exists$x(lamb(x))).
} \\[1ex]
(presumably trying to say ``Mary had some $x$ such that $x$ is a lamb'') the solver replies: \\[1ex]
{\tt
  \Indent{2} Input error on line 32:   had(Mary, SOME x lamb(x)). \\
  \Indent{2} Type mismatch with argument of had \\[1ex]
  \Indent{2} Detailed diagnostics: in the formula \\
  \Indent{4} had(Mary,SOME x lamb(x)) \\
  \Indent{2} the main operator "had" expects argument 2 to be of type animal \\
  \Indent{2} but argument 2 is \\
  \Indent{4} SOME x lamb(x) \\
  \Indent{2} which is of type bool.
}

\noindent
It would go on to suggest possible causes of this kind of error, advising for instance to check for misplaced parentheses and wrong names (not guilty in this case), and would also write out the parse tree of the formula as far as the parser was able to get before raising an exception. This kind of detail in error messages is an important feature, but such verbosity can become irritating so a future version of the site will place detailed drill-down under user control.

Semantic errors are harder to classify and harder to deal with. There are no ``canned'' solutions written in, so if the encoded problem gets past the parser and type checker, all the solver can do is search for solutions and report what it finds. Hence the only symptom of misunderstanding on the semantic level is an unexpected solution or (more often) no solution. This is the case whether the error is due to basic misunderstanding of semantics, such as confusion between the conditional and the biconditional, or whether it is a matter of problem representation---using logic correctly to say the wrong thing.

The case in which there is no solution is common, and of course the solver's response ``No solution found'' provides the user with a minimum of information. Techniques for making the feedback more informative include commenting out lines of the encoding and re-running to see whether solutions exist. This can sometimes be effective in pinpointing incorrectly expressed constraints, though it is laborious and the results are not always helpful. A diagnosis tool designed to help automate the process of isolating incorrect constraints in cases where the encoded problem globally has no solution is currently under development: a prototype exists, but has not yet been incorporated into the website, so there are no results yet concerning its usefulness in practice.

\section{Site usage}

\begin{figure}[t]
\centering
\includegraphics[width=120mm]{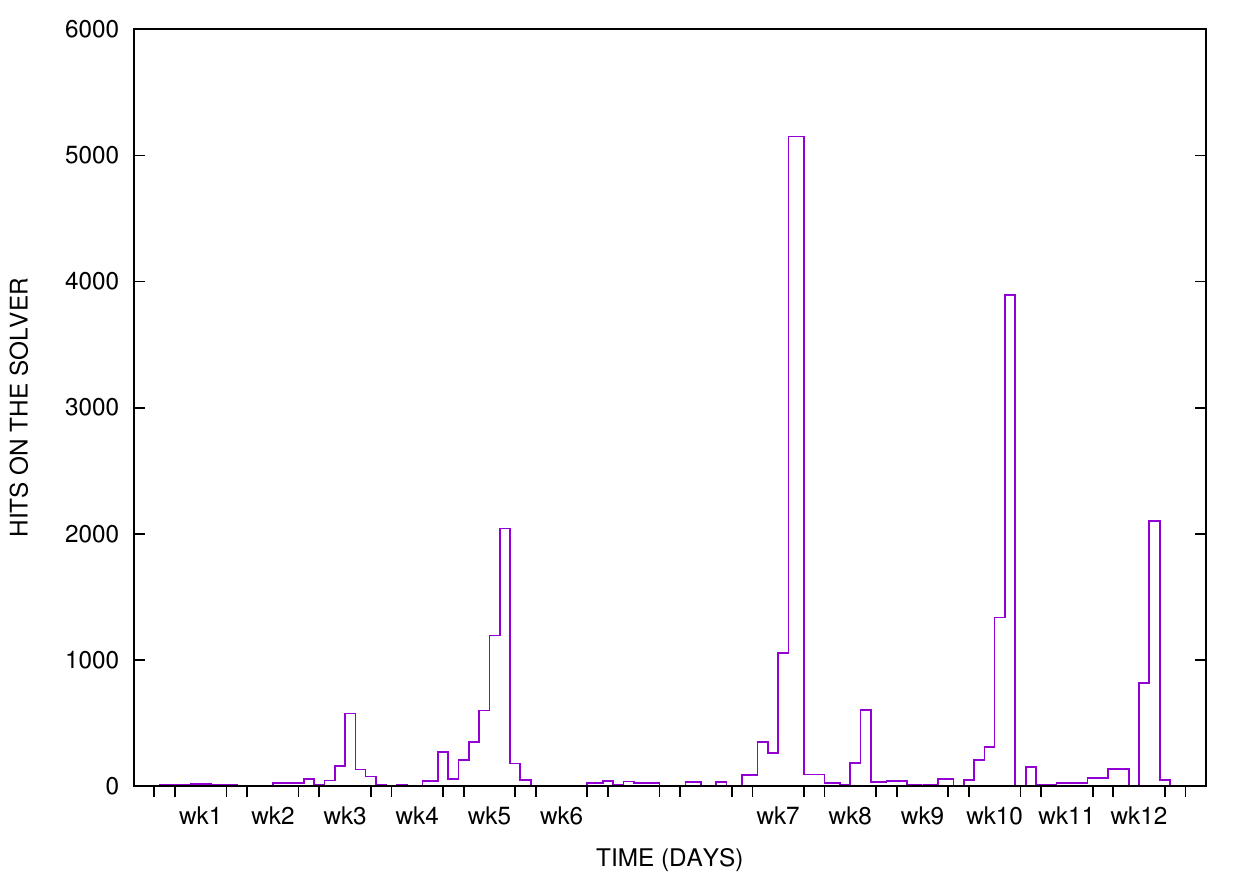}
\caption{\label{fig:byDays}Number of times the solver was run on each day over 12 weeks of a logic course}
\end{figure}

Logfiles produced by the scripts on the site can be mined for data on usage patterns, and provide some insight into how students set about mastering the web-based tool and using it to solve problems of logic. At the simplest, aggregated statistics for the number of hits on the site allow us to observe class behaviour. Figure \ref{fig:byDays} shows the number of times the solver was run each day by a class of around 50 undergraduates during a 12-week semester in 2013. Note that there was a 2-week break between weeks 6 and 7.

Students had a piece of homework to do each week, and had to hand it in for assessment before midnight each Friday evening. These assignments in weeks 5, 7, 10 and 12 consisted of problems to be solved using \emph{Logic for Fun}, as can clearly be seen from the bursts of activity in those weeks. The assignment in week 8 was a problem concerning the semantics of some first order formulae, which the students were asked to compute on paper, using semantic tableaux, before comparing their answers with models of the same formulae produced by \emph{Logic for Fun}. The small peak in week 3 is associated with the point at which they were introduced to the site and asked to complete some easy exercises to familiarise themselves with it. The homework in weeks 3, 4, 6, 9 and 11 did not call for students to use the site.

\begin{figure}[t]
\centering
\includegraphics[width=120mm]{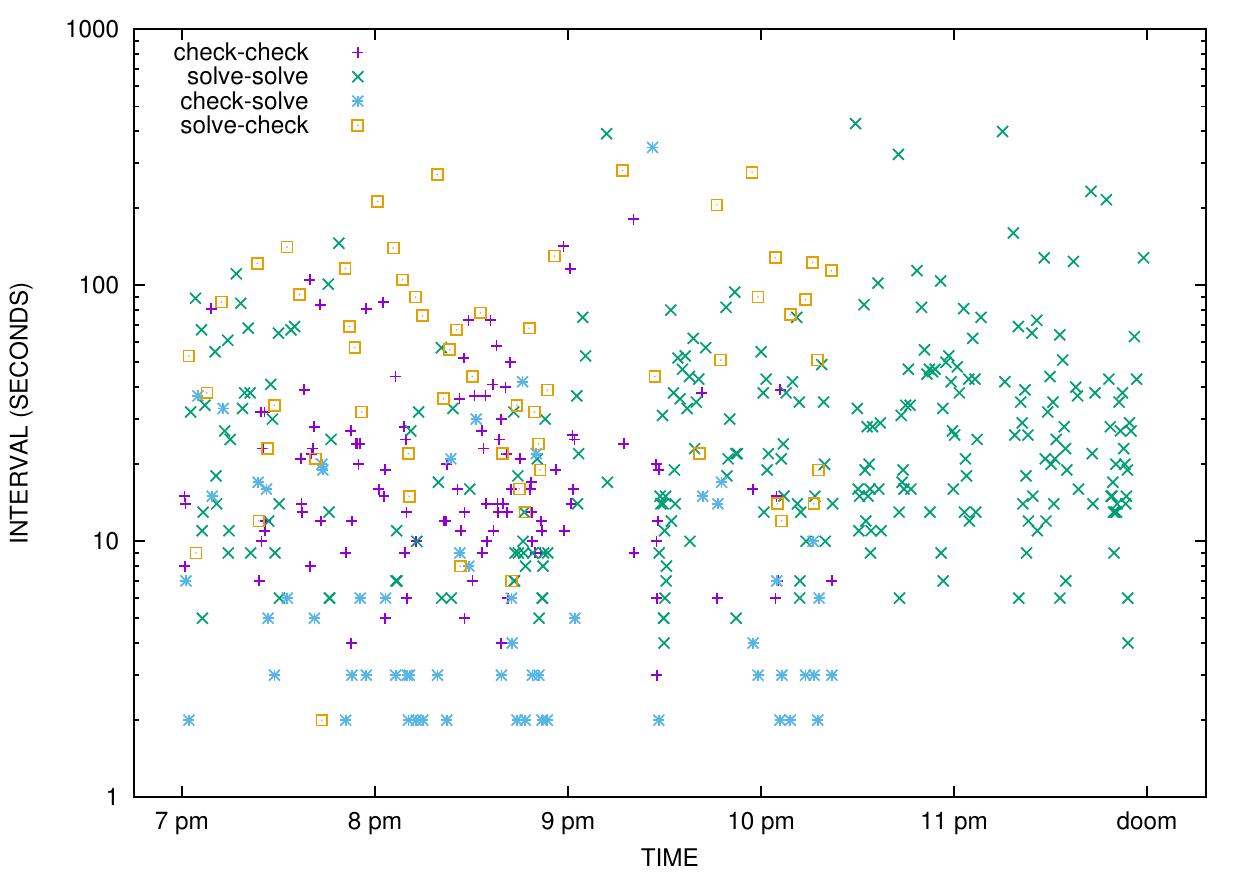}
\caption{\label{fig:fiveHours}Five hours of work by one student. Each point is a run of the solver, showing the interval since the previous run plotted against the time of day. Different shapes show whether the run was a syntax check or a ``solve'', and whether the previous run was a check or a solve.}
\end{figure}

The heaviest usage of the semester occurred on the Friday of week 7, when the solver was run on average almost 100 times per student. This represents an extraordinary amount of work by the class on what was a fairly modest piece of logic homework. The problem in question (see Appendix \ref{sec:example}) was designed to turn on the correct handling of quantifiers. Since the usage log from that period only records who ran the solver at what time, not the full text of what they sent to the solver, we have no way now of knowing what particular difficulties caused the class to spend so much more effort on this problem than on the others.

More detail is revealed by studying the work patterns of individual students. Some behaviours are quite striking: there are, for example, students who will run the solver 200 times or more on one problem, many of the runs being only seconds apart. Figure \ref{fig:fiveHours} shows an example of the activity of one such student over the five hours before the submission deadline. Note that this is a huge amount of work compared with the few minutes which students normally spend on a hand-written formalisation exercise. At no point during the five hours does this student pause for much more than 5 minutes. Most of the runs which occur within 10 seconds of the previous run are cases in which a syntax check is immediately followed by a ``solve'', presumably because there was no syntax error. We see some patterns in the record: for instance, at some point (a little before 10:30 pm) this student abandons explicit syntax checks and simply uses the ``Solve'' button. It is unclear why.

\section{Current and future work}

\emph{Logic for Fun} was completely re-scripted in 2013--14, partly because the look and feel of the old site dated from another millennium, partly to extend its functionality in significant ways, and partly to have a version built with modern tools which would be easy to maintain. The new site is scripted entirely in Python, though the solver behind it is still the original, written in C some 20 years ago. The beta version of the new site \cite{LFFnew} is now freely available to all web users, in contrast to the old site which required them to have accounts and to pay fees. This is in line with the contemporary expansion of free access to educational tools.

The biggest enhancement from the user's viewpoint is a facility to save and reload work, allowing it to be carried over easily from one session to another. Students who join a class (called a ``group'' on the site) can also submit their work to the group manager for feedback or assessment. Low-level improvements, such as organising the display so that the natural language version of a puzzle can be in the same browser window as the student's logical encoding of it, also do much to enhance the user experience.

There is an ambitious plan for continued upgrading of the site. Tools currently under development include the diagnosis assistant already noted above, for use when a problem as written by the user has no solution. This tool allows the user to see minimal unsatisfiable cores of the problem, either on the high (first order) level or at lower levels from the clausified and compiled versions of the logic. It can also present approximate solutions---assignments of values which satisfy \emph{almost} all of the constraints. A prototype of this tool looks promising, but incorporating it into the website raises issues for the user interface which have yet to be satisfactorily resolved. It is also planned that the system will maintain a detailed database of user activity, recording every character of text sent to the solver. This information will be used in a project aimed at deeper understanding of the logic learning process.

An important piece of future work, to be conducted when the new site is stable, is an effectiveness study. It is obvious that doing formalisation exercises online rather than on paper causes students to put much more effort into getting their answers right, but measuring the extent to which they learn more as a result is much harder.

\subparagraph*{Acknowledgements}

I wish to thank Matt Gray, Kahlil Hodgson, Daniel Pekevski, Nathan Ryan and Wayne Tsai for help in scripting \emph{Logic for Fun}, and the many logic students over the last 15 years who have helped by testing it to destruction.

\appendix

\section{Example: the homework in week 7}
\label{sec:example}

\begin{center}
\parbox[t]{100mm}{
{\sf
{\large\bf Logic Games} \\[1ex]
Our annual logic competition came to a final showdown between five teams: the Aces, Buccaneers, Cougars, Demons and Eagles. The contest was a round robin, each team meeting each other team once. At the end, the judges announced:
\begin{itemize}
\item Every team won at least once, and some team won all its games.
\item The Buccaneers beat only the Cougars.
\item Exactly one match was drawn, and it didn't involve the Cougars.
\item The Aces defeated every team that the Eagles defeated, but they didn't defeat the Demons.
\item Not every team that defeated the Aces defeated every team that the Aces defeated.
\end{itemize}
With that, they left us to work out the full set of results. Well?
}}
\end{center}

\bibliography{L4F}

\end{document}